\documentclass[conference]{IEEEtran}

\usepackage{amsmath,amssymb,amsfonts}
\usepackage{algorithmic}
\usepackage[ruled, linesnumbered]{algorithm2e}

\usepackage{graphicx}
\usepackage{textcomp}
\usepackage{xcolor}
\usepackage{booktabs}
\usepackage{multirow}
\usepackage{array}
\usepackage{caption}
\usepackage{balance}

\begin{document}

\title{Demystifying Gate-Level Localization of\\ RTL Trojans}

\author{
\IEEEauthorblockN{Navid Nader Tehrani, Azadeh Davoodi}
\IEEEauthorblockA{\textit{University of Wisconsin-Madison} \\
Madison, USA \\
\{nadertehrani, adavoodi\}@wisc.edu}
\and
\IEEEauthorblockN{Rasit Onur Topaloglu}
\IEEEauthorblockA{\textit{Marist University }\\
Poughkeepsie, NY, USA \\
rasit.topaloglu@marist.edu}
}

\maketitle

\begin{abstract}
Hardware Trojans are malicious modifications that compromise functionality or leak sensitive data. They pose a severe threat, particularly when inserted at the Register Transfer Level (RTL). After synthesis, these Trojans are often concealed by optimizations in gate-level netlists. 
Recent efforts, including the ICCAD 2025 contest, emphasize golden-chip-free detection using machine learning (ML) on labeled netlists. In this work, we show that RTL Trojans exhibit stable structural and signal-flow patterns post-synthesis, enabling effective detection through targeted heuristics rather than generic ML feature learning. We propose LoRD, a lightweight heuristic-based approach that exploits these distinctive subgraph signatures, achieving near-perfect detection and localization on the contest testcases. Compared to a transformer-based ML baseline and top five teams, LoRD achieves on-average a score of 2.957 (out of 3) for Trojan-implanted designs without the data and tuning overhead.
\end{abstract}

\section{Introduction}
The globalization of the integrated circuit (IC) supply chain has made hardware security increasingly difficult to guarantee. Today’s designs are handled by numerous third-party IP vendors, design houses, and foundries, any of which may be untrusted. This opens the door to malicious modifications of the circuitry, commonly referred to as Hardware Trojans \cite{HT_intro_2010, HTSurvey_2010}, which can stealthily alter functionality, degrade reliability, or leak sensitive information while remaining dormant under normal operating conditions. 

Hardware Trojans can be introduced at multiple points in the design and fabrication flow. Among them, Register Transfer Level (RTL) Trojans may be considered one of the most challenging to detect at lower levels. In this attack model, a Trojan module is maliciously inserted in the design at the RTL stage. As the design is synthesized and translated to a flattened gate-level netlist, detection and localization of the Trojan becomes extremely challenging. This is because commercial synthesis tools make the Trojan logic to blend seamlessly with the rest of the netlist at gate-level. Synthesis optimizations such as logic restructuring, resource sharing, retiming, and technology mapping can significantly distort both the functional and Trojan logic, so that an RTL Trojan may correspond to many structurally different gate-level implementations. 

Detecting RTL Trojans at gate-level has been the focus of the recent ICCAD 2025 contest (Problem A) which was organized by Cadence Design Systems \cite{ICCAD2025-contest}. 
Traditional defenses rely on comparing side-channel measurements (e.g., power \cite{SCA_power_TC13}, delay \cite{SCA_delay_HOST15, SCA_delay_ICCAD16}, or temperature \cite{SCA_tmperature_ICCAD13}) against a trusted `golden chip', which is often difficult and expensive to obtain for modern SoCs.

The contest therefore focuses on golden-chip-free detection: participants must identify RTL-inserted Trojans directly on synthesized gate-level netlists. The emphasis is on identifying Trojans based on the expected functionality of the RTL Trojan at gate-level. Use of machine learning (ML) techniques to learn Trojan patterns from data is heavily encouraged in the contest by providing labeled netlists, for both Trojan-implanted and Trojan-free cases.

There has been a large body of work exploring the use of ML for gate-level Trojan detection. Existing techniques often  extract a fixed set of handcrafted features for each gate or net—such as fan-in/fan-out counts, gate type, local path depth, controllability/observability measures, and loop or reconvergence indicators—and feed them to conventional classifiers including multilayer perceptrons  \cite{MLP11Features_IOLTS17},  support vector machines \cite{SVM5Features_IOLTS16, SVM_ICSIP21}, and deep learning models \cite{Yu2021TrojanDetection}. More recent methods utilize graph neural networks to learn node embeddings and capture higher-order structural information \cite{TrojanSAINT_ISCAS23, HW2VEC_2021, GNN4HT_2023}. Most recently, transformers and Large Language Models have also been used for gate-level Trojan detection \cite{Latibari2025Transformers, NtNDet_2025, chen2024trojanformer}. 

In this work, we show  detection of RTL Trojans at gate-level heavily relies on identifying distinctive structural and signal-flow patterns after synthesis, e.g., specific chains of flip-flops, characteristic trigger trees, and recognizable interactions with bus signals and status flags. These patterns are largely stable across synthesis optimizations and libraries, and they manifest as relatively small, well-structured subgraphs embedded in an otherwise large but regular design. As a result, these Trojans may not particularly benefit from generic ML feature learning 
which have additional hyperparameter tuning requirements.

To validate this insight, we implemented a Bidirectional Encoder Representations from
Transformers (BERT)-based ML baseline \cite{BERT_arXiv18}, where localized netlist contexts are tokenized and classified if they contain a Trojan. This baseline was trained using the labeled gate-level public netlists provided by the contest. 
When evaluated using their hidden testcases, the BERT-based scheme only achieves a moderate  localization performance in Trojan-implanted designs. In contrast, `LoRD', our proposed scheme, attains an almost-perfect score under the contest metrics for all hidden cases. 

The summary of our contributions is listed below:
\begin{itemize}
    \item We demonstrate that RTL Trojans exhibit stable structural and signal-flow patterns after synthesis, making them identifiable through targeted heuristics rather than generic ML feature learning.
    \item We introduce LoRD, a lightweight detection scheme that exploits distinctive subgraph signatures of Trojans, avoiding the data and tuning overhead of ML models while achieving superior accuracy.
    \item Through comparison with a BERT-based ML baseline trained on contest data and top five teams, LoRD attains near-perfect detection and localization scores on hidden test cases, for both Trojan-implanted and Trojan-free.
    \item LoRD is computationally light, and is able to process the entire 60 hidden test cases in under a minute which is significantly below BERT's training and inference time.
\end{itemize}

\section{Problem Formulation}
Given a set of RTL Trojans, and a flattened gate-level netlist, our goal is to detect if the netlist includes any of the Trojans. Additionally, if a Trojan is detected, our goal is to localize it by predicting all gates in the netlist which belong to the Trojan.

In this work, we follow the same setup as the recent ICCAD 2025 Contest Problem A \cite{ICCAD2025-contest}, organized by Cadence Design Systems. The contest targets detection and localization of ten RTL Trojan types at gate-level. These Trojan types are placed into four groups based on functionality (i.e., information leakage, trigger events, control flow manipulation, and selective logic modification). These are summarized in Columns 1 and 2 of Table \ref{tab:trojan_detection_results}. Detection and localization of the RTL Trojan in a given netlist is challenging because the synthesis options and used library may be unknown. Additionally, different variations of an RTL Trojan may be used, for example by changing constant values, bus widths, etc. We explain more details in the upcoming section as we discuss our methods.

\section{Our Methods}

We first present details of LoRD, and then present an alternative transformer-based ML approach based on BERT \cite{BERT_arXiv18}.

\subsection{Trojan Detection \& Localization with LoRD}

Given a synthesized gate-level netlist, LoRD applies a suite of carefully crafted yet lightweight heuristics to the gate-level design, with one heuristic per Trojan type\footnote{We found a single  heuristic to be sufficient for detecting   Trojans~8 and~9 because they are quite similar in the structure of their RTL codes.}. Each heuristic returns a Boolean value indicating if its corresponding Trojan $t$ has been detected. Additionally, upon detecting Trojan $t$, the heuristic outputs a set of gates, $\mathcal{L}_t$, representing candidates that match the Trojan’s structure at gate-level. Detection and localization of a Trojan are based on the topology and signal-flow cues identified specifically for each Trojan group such as chain structure, logic function on critical arcs, fan-in/out patterns, etc. which are quite different than handcrafted numerical features used by many ML-based schemes. 
In case several heuristics indicate detection of Trojans, the one with the greatest $\mathcal{L}_t$ size determines the final result.  

LoRD  utilizes discriminative cues based on topology and signal flow in the netlist to detect and localize each group of RTL Trojans at gate-level. Next, we provide a detailed walk-through of LoRD's procedure for each group. We focus on \textit{one representative Trojan per group} due to limited space. We note that the functionalities of the Trojans in each group are often quite similar to each other.

\subsubsection{Group~1 Trojans: Information Leakage} 
\begin{figure}[t]
    \centering
\vspace{-3mm}    \includegraphics[width=1\linewidth]{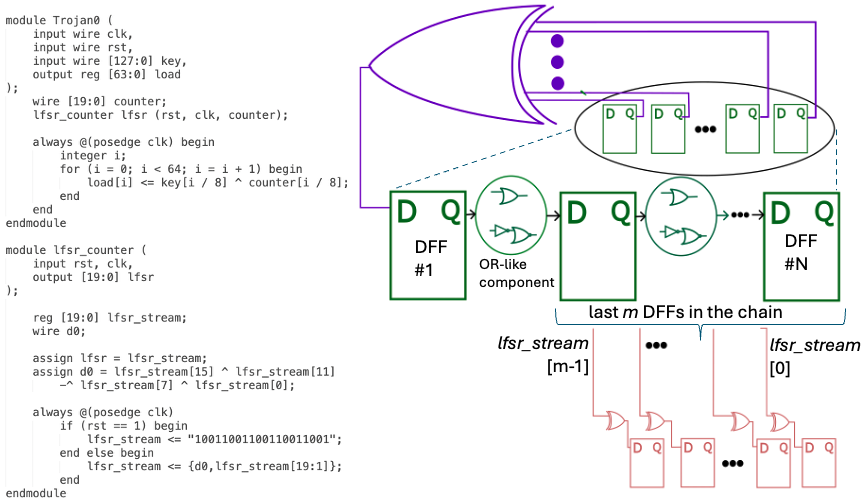}
\vspace{-5mm}
    \caption{Overview of RTL of Trojan T0 (Group 1) and our clues for gate-level detection and localization.}
    \label{fig:T0}
\end{figure}

This group aims to leak sensitive data such as encryption keys. 
Figure \ref{fig:T0} shows the RTL description of T0 and its gate-level clues. This Trojan instantiates an LFSR/shift-register–like chain of DFFs of size $N$. The DFF at each stage is driven by a small OR-like component  from earlier stages, producing a pseudo-random stream (denoted by \texttt{lfsr\_stream}). The role of the OR-like component is to initialize the LFSR with a seed using the \texttt{rst} signal. 
The payload of the Trojan exposes internal secret bits by XORing the last $m$ stages with the \texttt{key} data and latching those in additional  DFFs to drive the \texttt{load} pins.

\noindent\textit{Graph model and primitives:} We convert the netlist to a directed graph $G=(V,E)$ with each gate represented as a node, and a directed edge connecting each driver pin to its sink pin.
We classify `OR-like' components as a grouping of gates which implement a two-input OR functionality. For the set of library gates provided by the contest, this set is given by: \{\texttt{OR}, \texttt{NOT connected to NOR}\}. A \emph{chain} edge $u\to v$ (from DFF $u$ to DFF $v$) exists if $u.\texttt{Q}$ reaches $v.\texttt{D}$ through one OR-like component and no other reconvergent logic.

\noindent\textit{Procedure:} Algorithm 1 provides the psuedo-code to detect and localize T0. We first search for the longest LFSR-like chain in $G$ (lines 1-8). In line 3, \textit{StartCandidate$(v)$} indicates if DFF $v$ qualifies as start of an LFSR chain. This happens if 
(1) $v.\texttt{D}$ is feeding an OR-like component followed by a DFF, 
and (2) $v$ isn't driven by an OR-like component. Next, in line 11, the feedback loop of the LFSR is identified (shown in purple). Finally, in lines 13-19, the leaky outputs which reveal the key are identified (shown in red). The Trojan gates are compromised of the LFSR chain, its feedback loop and the leaky outputs which are generated in lines 22-24. 

\noindent\textit{Main Observations:} Algorithm 1 only relies on detecting the \textit{structure} of the LFSR and payload. No assumptions are made about the signal names and their length, the LFSR seed and  chain length (except a threshold $\tau$ which we set to 4).

\begin{algorithm}[t]
\caption{Group 1 (T0): Detection \& Localization}
\small
\begin{algorithmic}[1]

\REQUIRE Netlist graph $G=(V,E)$

\STATE \textit{Step 1: Identify the LFSR chain}

\FOR{each DFF $v \in V$}
    \IF{StartCandidate$(v)$}
        \STATE $C \gets$ GrowChainFrom$(v)$
        \IF{$|C|$ is longest so far}
            \STATE $C^\star \gets C$
            \STATE $v^\star \gets v$
        \ENDIF
    \ENDIF
\ENDFOR

\STATE \textit{Step 2: Identify the feedback loop of the LFSR}

\STATE Perform backward DFS from $v^\star$
until reaching any component in $C^\star$

\STATE \textit{Step 3: Find leaky outputs}

\FOR{$i=N$ down to $1$}
    \IF{$C^\star[i]$ connects to an XOR through Q and the XOR is followed by a DFF}
        \STATE Mark the DFF and XOR as Trojan
    \ELSE
        \STATE \textbf{break}
    \ENDIF
\ENDFOR

\STATE \textit{Step 4: Output Trojan gates}

\IF{length$(C^\star) > \tau$}
    \STATE Return \textbf{true} and all gates in
    $C^\star$, the feedback loop, and the leaky outputs
\ENDIF

\end{algorithmic}
\end{algorithm}

\subsubsection{Group 2: Trigger Events}
\begin{figure}[t]
    \centering
\vspace{-3mm}    \includegraphics[width=1\linewidth]{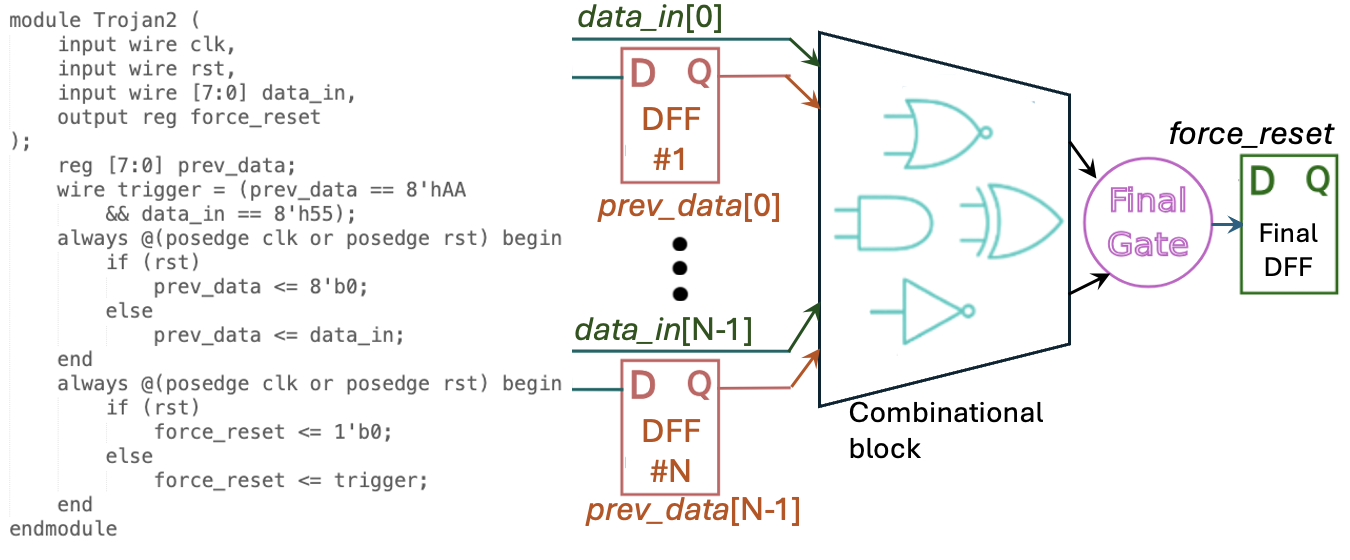}\vspace{-2mm}
\caption
 {Overview of RTL of Trojan T2 (Group 2) and our clues for gate-level detection and localization.}
    \label{fig:T2}\vspace{-4mm}
\end{figure}

This group of RTL Trojans modifies control-related signals (e.g., reset or enable) under
specific input conditions, thereby causing functional misbehavior or system interruption.
Figure~\ref{fig:T2} shows the RTL description of T2 from this group. The Trojan continuously
stores the previous input vector \texttt{prev\_data} and compares it with the current
\texttt{data\_in}. When the two consecutive values form a specific sequence (e.g. \texttt{8'hAA} followed
by \texttt{8'h55}), \texttt{trigger} is asserted, making \texttt{force\_reset} high.
Overall, T2 injects a conditional reset pulse that activates under a rare input pattern.

\noindent\textit{Graph model and primitives:}
In the gate-level netlist, T2's structure appears as a group of DFFs capturing consecutive
data bits (\texttt{prev\_data[7:0]}) with a combinational comparison block feeding a single
output DFF (\texttt{force\_reset}). The DFFs share a common input bus
(\texttt{data\_in[7:0]}), forming a synchronized data-register
bank. The trigger condition corresponds to an equality comparator tree between the stored
data and constant nets (e.g., \texttt{8'hAA}, \texttt{8'h55}).

\noindent\textit{Procedure:}
Algorithm~\ref{alg:t2} summarizes the steps for detection and localization of T2. It relies on three traits:
(1) a bank of $N$ parallel DFFs (\texttt{prev\_data[7:0]}) with shared bus inputs and outputs;
(2) a forward connection from this bank into a single DFF that controls a reset-like signal;
and (3) a closed backward path from that control DFF to the same input bus through
comparison logic. 

These features---collectively identifying a `triggered control injection'
pattern---are rare in Trojan-free designs and  consistent with the rare-condition activation style
of Group~2 Trojans.

\noindent{\textit{Main Observations:}} Algorithm 2 relies on identifying the Trojan based on few assumptions which are independent of specific gate-level implementation styles. These include identifying the input and output bus lines without making any assumption about the width of the bus (except having a minimum threshold $\tau$ which we set to 6 in our implementation). We do assume lines within the same bus to have a common name but do not make assumption about what the name may be in the RTL Trojan module. Additionally, we do not make assumptions about how the combinational logic implementing comparison of previous and current data may be implemented. These are detected via tracing signal flow in a combinational logic block feeding a single DFF. 

\begin{algorithm}[t]
\caption{Group 2 (T2): Detection \& Localization}
\label{alg:t2}
\small

\KwIn{Netlist graph $G=(V,E)$}

\textbf{Step 1: Identify register banks}\;

Find a set of DFFs
$\mathcal{D}=\{d_1,\ldots,d_n\}$ with $n>\tau$
such that all $d_i$ share a common input-bus prefix
(\texttt{nx[i]}) and a common output-bus prefix
(\texttt{ny[i]})\;

\textbf{Step 2: Forward traversal}\;

Select any $d \in \mathcal{D}$ and perform a forward DFS in $G$
to locate a reachable DFF\;

Denote the discovered DFF by $d_{\mathrm{final}}$\;

\textbf{Step 3: Backward traversal}\;

Starting from $d_{\mathrm{final}}$, perform a bounded backward BFS
following driver edges until reaching the input nets of
$\mathcal{D}$ (the bus \texttt{nx[i]})\;

Collect all visited gates and DFFs into a set $T$\;

\textbf{Step 4: Output Trojan gates}\;

\If{$\mathcal{D} \subseteq T$}{
    Return \textbf{true} and output all nodes in $T$
    as Trojan gates\;
}

\end{algorithm}

\subsubsection{Group 3: Control Flow Manipulation}
\begin{figure}[t]
    \centering 
    \includegraphics[width=1\linewidth]{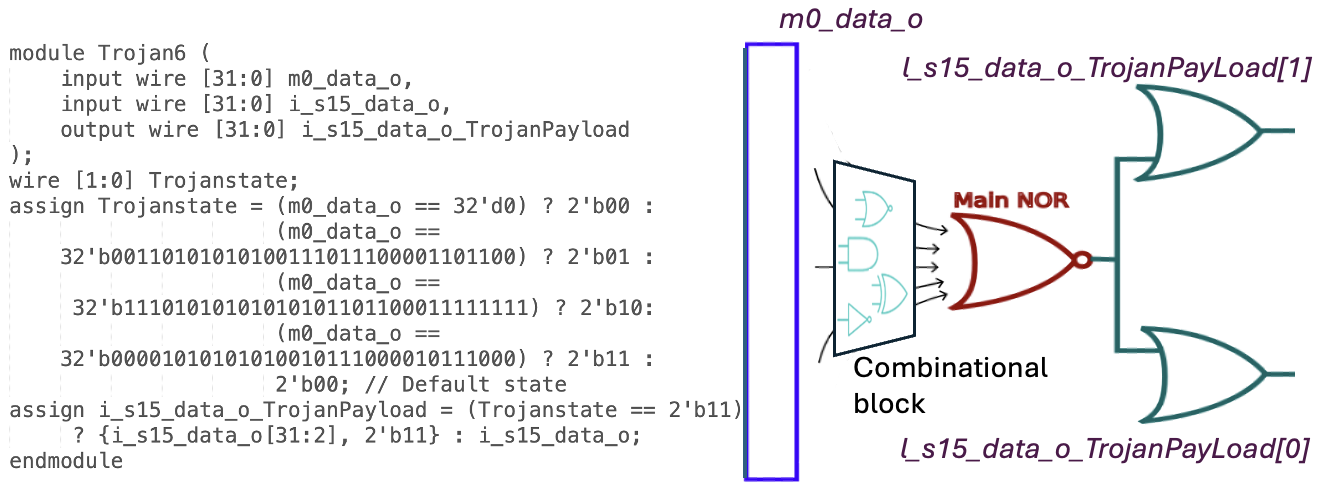}
    \caption{Overview of RTL of Trojan T6 (Group 3) and our clues for gate-level detection and localization.}
    \label{fig:T6}\vspace{-4mm}
\end{figure}

This group of RTL Trojans targets control-flow by injecting a malicious control pattern via selectively
overwriting a few bits of a bus (e.g., a program counter). The overwriting happens when a hidden multi-bit condition on an internal
data bus is satisfied. Figure~\ref{fig:T6} shows the RTL description of
\texttt{T6}. The \texttt{m0\_data\_o} bus (32 bits in the RTL) is decoded into a 2-bit
signal \texttt{Trojanstate}, which encodes several rare constants. When
\texttt{Trojanstate} becomes \texttt{2'b11}, the two
least significant bits of \texttt{i\_s15\_data\_o} are set which changes
\texttt{i\_s15\_data\_o\_TrojanPayload} 
to an unintended state or address.

\noindent\textit{Graph model and primitives:}
By examining the labeled gate-level netlists provided by the contest for this Trojan, we observed the condition on \texttt{Trojanstate} is synthesized as a small
decode network whose output is merged with two bits of the target bus through
OR gates. In the gate-level netlist, the core of this network is a single
NOR gate whose output fans out exclusively to two OR gates (or OR-like components) that drive two bits
of the same bus. Upstream of this NOR gate lies a cone of comparators (combinational logic) that
checks equality of \texttt{m0\_data\_o} against fixed constants. 

\vspace{1mm}
\noindent\textit{Procedure:}
Algorithm~\ref{alg:t6} summarizes the heuristic detection of this Trojan. We exploit two
structural properties:
(1) the presence of a \texttt{main NOR} gate whose only fanout is to two OR gates,
and (2) the fact that the entire fan-in cone of this NOR terminates at a bus net (nets whose names contain ``\texttt{[}'', e.g.,
\texttt{nx[31]}).

\begin{algorithm}[t]
\caption{Group 3 (T6): Detection \& Localization}
\label{alg:t6}
\small

\KwIn{Netlist graph $G=(V,E)$}

\textbf{Step 1: Locate main NOR gates}\;

Identify candidate \texttt{main NOR} gates by scanning all NOR gates
$v \in V$ and selecting those whose fanout consists of exactly two
OR-like components, denoted by $o_1$ and $o_2$\;

\textbf{Step 2: Backward traversal}\;

For each candidate \texttt{main NOR} gate $v$, perform a backward BFS
starting from $v$ and following driver edges until reaching nets whose
names contain ``\texttt{[}'' (e.g., bus wires such as \texttt{nx[i]})\;

Collect all visited gates, including $v$, into a set
$T_{\mathrm{BFS}}$\;

\textbf{Step 3: Mark Trojan cone}\;

Form the candidate Trojan set

\[
T = T_{\mathrm{BFS}} \cup \{o_1,o_2\},
\]

where $o_1$ and $o_2$ are the OR-like components driven by the
\texttt{main NOR}\;

\textbf{Step 4: Output Trojan gates}\;

\If{a candidate \texttt{main NOR} $v$ is found and $|T| > \tau$}{
    Return \textbf{true} and output $T$ as the set of Trojan gates\;
}

\end{algorithm}

\noindent\textit{Main Observations:}
Trojan~6 exhibits a distinctive topology: a single decode signal (the NOR
output) controls exactly two bits of a bus through OR gates, and the entire
decode network is driven solely by comparators on a different bus.

Benign control logic rarely dedicates a separate NOR--to--two-OR chain to
override just two bus bits based purely on such a deep equality-check cone.
By explicitly searching for this pattern and enforcing a size threshold
($|T| > \tau$ which we set to 24 in our implementation), we show our procedure is able to reliably isolate the malicious control-flow manipulation introduced by this Trojan while avoiding most normal logic. Additionally, our procedure 
assumes lines within the same bus to have a common name
but do not make assumption about what the name may be
in the RTL Trojan module. The width of the bus could be any constant higher than a threshold which we set to 8 bits in our implementation.

\subsubsection{Group 4: Selective Logic Modification}

\begin{figure}[t]
    \centering
    
    \vspace{-2mm}
    \includegraphics[width=1\linewidth]{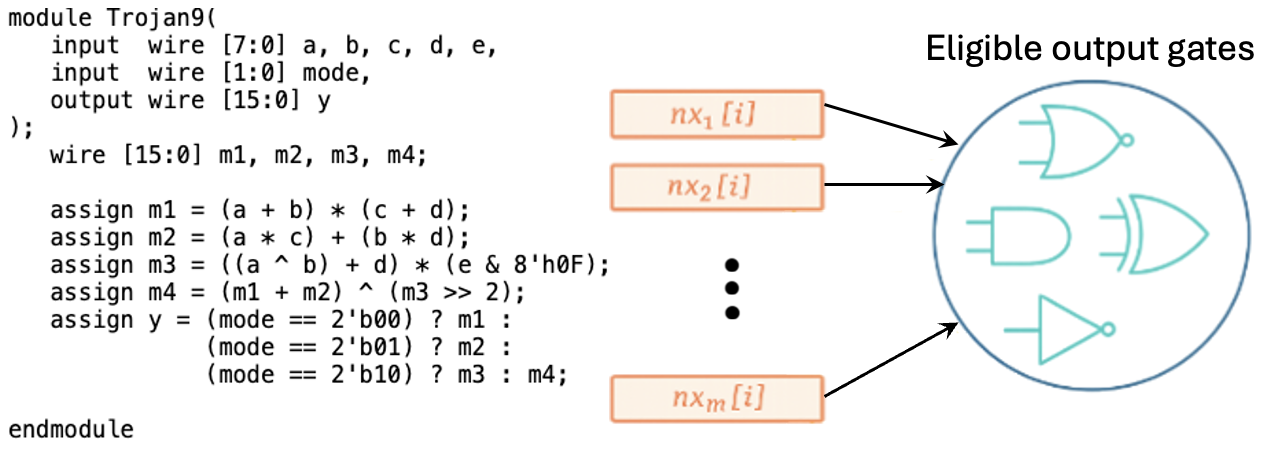}
    \caption{Overview of the RTL and  clues for gate-level detection and localization of Trojan T9 (Group 4).}
    \label{fig:T89}\vspace{-4mm}
\end{figure}

This group of RTL Trojans conditionally changes logic outputs. They are implemented as small arithmetic {engines} whose outputs can be switched among several internally–generated expressions. 
Figure~\ref{fig:T89} illustrates the RTL and the detection scheme for Trojans T8 and T9. As can be seen from the RTL (for T9), the Trojan module takes multiple 8-bit input words \texttt{a,b,c,d,e} and computes several 16-bit intermediate results (\texttt{m1--m4}). A \texttt{mode} signal  selects which intermediate word drives the 16-bit output \texttt{y}. Some of these expressions are mathematically related (e.g., distributive rewrites), while others inject masked or shifted versions of the data (e.g., bit-masking with \texttt{16'h00FF} or \texttt{8'h0F}, or right shifts). Thus, under particular modes the Trojan quietly substitutes a modified arithmetic function for the original one, changing only selected bits of \texttt{y} while reusing the same inputs and apparent interface.

\vspace{1mm}
\noindent\textit{Graph model and primitives:}
At the gate level, the datapath of this Trojan appears as a dense combinational cone rooted at the bits of the output bus \texttt{y[i]}. Each root gate has a single fanout to a net whose name contains an index, e.g., \texttt{nx\_out[i]}. Tracing backwards from such a gate, its fan-in cone reaches many other bus bits with different base names (e.g., \texttt{a[i]}, \texttt{b[i]}, \texttt{c[i]}, \texttt{d[i]}, \texttt{e[i]}), as illustrated in Figure~\ref{fig:T89}. Our  scheme exploits the textual naming convention of bus nets: all bits of a word share a common prefix and differ only in the index inside the brackets. For a net named \texttt{nx\_k[i]}, the word name is defined as the substring before the bracket, \texttt{nx\_k}. Our goal is to find output-bit cones whose logic mixes a large number (higher than  $\tau$=4) of distinct word names.

\begin{algorithm}[t]
\caption{Group 4: Detection \& Localization (T8 \& T9)}
\label{alg:t89}
\small

\KwIn{Netlist graph $G=(V,E)$}

\textbf{Step 1: Find candidate output gates}\;

Let $\mathcal{O}$ be the set of combinational gates whose output
net name contains ``['' (i.e., they drive a bus bit such as
\texttt{ny[i]})\;

\ForEach{$g \in \mathcal{O}$}{

    Initialize a set of word names $W \gets \emptyset$\;

    Initialize a gate set $T \gets \emptyset$\;

    \textbf{Step 2: Backward traversal}\;

    Perform a backward BFS through combinational logic starting from $g$\;

    \ForEach{visited edge $(u \rightarrow v)$ with net name $n$}{

        \If{$n$ contains ``[''}{

            Extract the word name $w$ as the substring preceding ``[''\;

            Add $w$ to $W$\;

            Do not traverse beyond net $n$\;
        }

        Add gate $u$ to $T$\;
    }

    \textbf{Step 3: Output Trojan gates}\;

    \If{$|W| > \tau$}{
        Mark $g$ and all gates in $T$ as Trojan gates\;
    }
}

\If{$|W| \neq 0$}{
    Return \textbf{true} and output $T$ as the Trojan gate set\;
}

\end{algorithm}
\noindent\textit{Procedure:}
Algorithm~\ref{alg:t89} lists the procedure for Trojans~T8 and~T9.

\noindent\textit{Main Observations:} In Algorithm 4, each bit of \texttt{y} is computed from arithmetic expressions that deliberately mix many different input words: additions and multiplications over \texttt{a,b,c,d,e}, followed by masking, shifting, and XORing with constants. Consequently, the gate-level fan-in cone of any \texttt{y[i]} contains leaf nets from at least five distinct word prefixes (\texttt{a[·]}, \texttt{b[·]}, \texttt{c[·]}, \texttt{d[·]}, \texttt{e[·]}). Typical functional logic in the design netlists rarely exhibits such heavy cross-word mixing for a single bus bit. By explicitly counting how many distinct word-level buses feed each candidate output gate, our scheme isolates precisely those cones that behave like the `selective arithmetic replacement' blocks in these Trojans. All thresholds $\tau$ are chosen empirically based on public contest netlists (as specified for each Algorithm) and then held fixed for all hidden test cases.

\subsection{BERT-based Trojan Identification}\label{sec:BERT}

As an alternative approach and to provide a  point of reference, we developed a deep learning model using  Bidirectional Encoder Representations from Transformers (BERT) \cite{BERT_arXiv18}. This is in part because the ICCAD 2025 contest inspired the use of machine learning by providing labeled gate-level netlists to use as training data. 

\begin{figure}[t]
    \centering
    \includegraphics[width=1\linewidth]{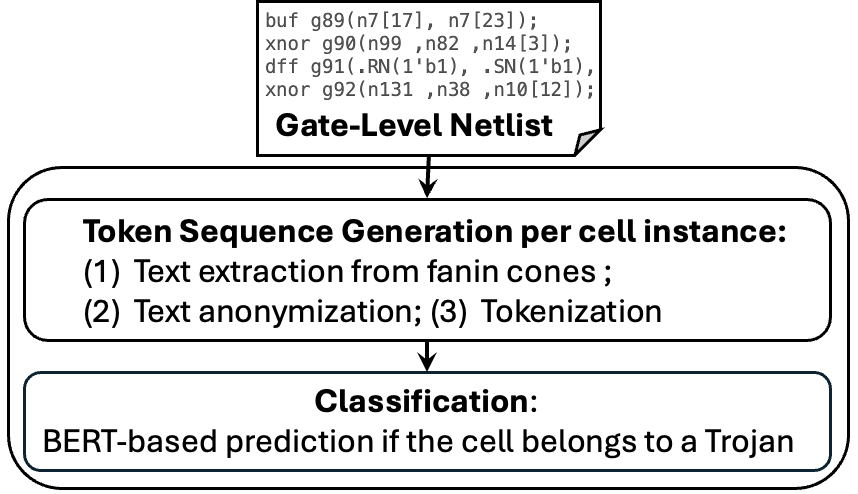}
    \caption{Overview of BERT-based Trojan localization}
    \label{fig:BERT}
\end{figure}

We first convert the gate-level netlist, by re-expressing the more-complex gates in terms of simple gates. Specifically, \texttt{NAND}, \texttt{NOR}, \texttt{XOR}, \texttt{XNOR}, and \texttt{BUF} in the  contest synthesis library, are replaced with functionally-equivalent variations using \texttt{AND}, \texttt{OR}, \texttt{NOT} in the netlist. 

This step ensures the subsequent BERT-based training and classification is independent of gate types in the library, emphasizing function-based identification of Trojan behavior. (A cell instance of a complex gate type would be classified as Trojan if each simple gate implementing it is also classified to be part of the Trojan.)

BERT is a powerful deep learning model with strong ability to understand context and relationships between tokens in a sequence. This makes it highly suitable for classifying sequences derived from circuit netlists, as it can capture long-range dependencies and subtle patterns that might indicate malicious modifications. The motivation behind selecting BERT compared to other large language models (such as LLaMA \cite{touvron2023llama}) is its relatively smaller size, requiring fewer samples for training and its encoder nature, which can be better in detection tasks. Additionally, BERT’s modular design makes it straightforward to
implement custom embedding strategies tailored specifically for circuit analysis.
The key insight behind BERT-based Trojan identification is that gate-level netlists exhibit patterns similar to natural language. They have syntax (how gates are connected), semantics (what functions circuits perform), and context (how components relate to each other). 

To maximize success rate per Trojan group, after converting the netlist, we train a separate BERT model for each group of Trojans for a total of four models. Figure \ref{fig:BERT} illustrates an overview of BERT-based Trojan identification for one group. Starting from a gate-level netlist as input, for each cell instance, a token sequence is generated which is fed to the BERT model to predict how likely the cell is to be part of a considered Trojan group. 
There are two main components: (1) classification with BERT-based model for each cell instance  in a considered netlist, and (2) training the BERT model. A cell is classified as Trojan if its probability of detection by BERT is higher than a specified threshold.



We used a BERT-based binary classifier wrapped in a custom PyTorch \texttt{nn.Module} which first builds token representations with a custom embedding layer. It combines default word and  positional embeddings, and a custom tree-based embedding \cite{tree_emb_Shiv_2019} into vectors of size 512. These embeddings are then fed to a HuggingFace \texttt{BertModel} instantiated from a custom \texttt{BertConfig} including 4 transformer encoding layers with 8 attention heads per layer. The model passes the BERT output through a final linear layer to produce logits over the two classes (Trojan vs. non-Trojan). 
Overall, our BERT model had about 25M parameters which is similar in size to the \texttt{bert-small} model \cite{turc2019wellread}. This choice of BERT was in part due to availability of limited training data by the contest.



\noindent\textit{Classification with BERT-based model:} For a given netlist, we classify each cell instance using the BERT model for each Trojan group. To classify a cell, we generate a token sequence for it which involves text extraction, anonymization, and tokenization, as listed in Fig. \ref{fig:BERT}. 

First, for text extraction, we identify the line in the netlist implementing a cell instance, providing its input and output signal names. We then identify all other lines in the netlist implementing the fan-in cone of the cell for six levels. Next the extracted lines are reordered topologically so each cell appears prior to its fan-outs. These lines are then concatenated into one expression which serves as the context,  capturing functional and structural relationships by encoding 
signal paths that might indicate Trojan behavior. 

Next, we apply anonymization to the generated context. To ensure the model learns generalizable patterns, we remove naming information from the extracted text. This  involves  replacing specific cell instance names with their  gate types. Inputs to the fan-in cone are  anonymized as `X'. Finally, the anonymized context is  represented by a token sequence which is fed to the BERT model. We developed our own tokenizer. The tokens are the gate types in the contest library and `X' representing input to a fanin-cone. Other options such as WordPiece tokenizer \cite{wordpiece_arXiv21} may be used.

\noindent\textit{BERT model training:}\label{BERT_classification_training}  The generated token sequence for a gate instance is fed to BERT for classification to  output a probability if the gate belongs to a Trojan. 
To train a BERT model, we used public gate-level netlists provided by the contest. The Trojan type inserted inside each gate-level netlist was identifiable from the RTL, and the Trojan gates inside each netlist were specified by the contest.  The contest also provided two netlists containing each Trojan type (for a total of 20), and 10 Trojan-free netlists which were all used for our training. 
The training samples were generated by generating a token sequence for each cell. For labeling, samples corresponding to a Trojan cell (which are known from the public set) are assigned a label of `1', while samples corresponding to normal gates (and Trojan-free designs) are labeled with `0'. Positive and negative samples are balanced in training each BERT model. Training uses AdamW optimizer with learning rate 2×10$^{-5}$, batch size 8, maximum sequence length of 512 tokens, and the number of epochs is different for each model, varying from 4 to 12. Evaluation is done using 60 \textit{different} hidden test cases that the contest provided.

\begin{table*}
 \centering
 \caption{Comparison of LoRD and BERT-based approach for the 40 hidden Trojan-implanted test cases. Maximum F1 is 1 and maximum Score is 3. The metrics in each row are averaged across the corresponding Trojan-implanted netlist IDs.}
 \begin{tabular}{c|c|cccc|ccc|c}
   \toprule
   \multirow{2}{*}{\textbf{Trojan Function}} &
\multirow{2}{*}{\textbf{Trojan ID}} &
\multicolumn{4}{c|}{\textbf{LoRD}} &
\multicolumn{3}{c|}{\textbf{BERT-based}} & {\textbf{Netlist ID}}\\
& & \textbf{Precision} & \textbf{Recall} & \textbf{F1} & \textbf{Score} &
\textbf{Precision} & \textbf{Recall} & \textbf{F1} &  \textbf{identified by LoRD} \\

   \midrule

   \multirow{2}{*}{1) Information Leakage} & T0   & 0.986 & 0.963 & 0.974 & 2.974 & 0.165 & 0.986 & 0.273 &  0, 1, 20, 33\\
                   & T4   & 0.960 & 0.906 & 0.929 & 2.929 & 0.129 & 0.736 & 0.207 &  8, 9, 24, 32 \\
\cmidrule(lr){1-10}
\multirow{2}{*}{2) Trigger Events} & T1   & 0.979 & 0.978 & 0.978 & 2.978 & 0.048 & 0.710 & 0.087 &  2, 3, 21, 37\\
                   & T2   & 1.000 & 0.970 & 0.985 & 2.985 & 0.052 & 0.622 & 0.091  & 4, 5, 6, 22, 30\\
\cmidrule(lr){1-10}
\multirow{2}{*}{3) Control Flow Manipulation} & T5   & 0.988 & 1.000 & 0.994 & 2.994 & 0.114 & 0.963 & 0.186 & 10, 11, 25, 38\\
                   & T6   & 0.847 & 1.000 & 0.910 & 2.910 & 0.030 & 0.960 & 0.057  & 12, 13, 26, 31\\
\cmidrule(lr){1-10}
\multirow{3}{*}{4) Selective Logic Modification} & T3   & 0.995 & 1.000 & 0.997 & 2.997 & 0.138 & 0.816 & 0.226 & 7, 23, 35\\
                   & T7   & 0.971 & 1.000 & 0.985 & 2.985 & 0.040 & 0.737 & 0.075  & 14, 15, 27, 39 \\
                   & T8,T9 & 0.963 & 0.908 & 0.911 & 2.911 & 0.675 & 0.912 & 0.688 & 16, 17, 18, 19, 28, 29, 34, 36 \\
   \midrule
   \textbf{} & \textbf{Average} & \textbf{0.965} & \textbf{0.963} & \textbf{0.957} & \textbf{2.957} & \textbf{0.204} & \textbf{0.831} & \textbf{0.254} \\
   \bottomrule
 \end{tabular}%

 \label{tab:trojan_detection_results}
\end{table*}

\section{Experimental Results}



\subsection{Setup and Evaluation Metrics}

\noindent\textit{Testcases:} 
The ICCAD 2025 contest (Problem A) provides ten RTL Trojan types in Verilog with Trojans T0 to T7 from Trust-Hub \cite{TrustHub1, TrustHub2} and  Trojans T8 and T9  from industrial datapath designs \cite{ICCAD2025-contest}. Flattened gate-level netlists were also provided, which according to the contest specifications were composed of eight primitive gate types, flipflops (DFF with posedge \texttt{clk}, negedge reset), wires, and constants values. 
Each netlist had at most one type of Trojan. 

Additionally, 60 hidden test cases (flattened gate-level netlists) were provided for final evaluation which included 40 Trojan -implanted netlists (netlist IDs 0 to 39) and 20 Trojan-free designs (netlist IDs 40 to 59). The contest released evaluation results at its Beta stage which revealed which netlists were Trojan-implanted for the hidden cases but did not specify which Trojan type was included in a Trojan-implanted netlist. 
Additionally, the contest released 30 public cases which we used to train our BERT models. 



\noindent\textit{Evaluation Metrics and Requirements:} We use the same evaluation criteria specified in the contest: A score of two is assigned per netlist if it is correctly identified as Trojan-free or Trojan-implanted. 

Additionally, in a Trojan-implanted netlist, this Score is summed with the F1 score calculated for the netlist. The F1 score is calculated as follows.
\begin{align*}
\mathrm{Precision} &= \frac{\mathrm{TP}}{\mathrm{TP}+\mathrm{FP}}\quad
\mathrm{Recall} = \frac{\mathrm{TP}}{\mathrm{TP}+\mathrm{FN}}\\
\mathrm{F1} &= 2\cdot\frac{\mathrm{Precision}\cdot\mathrm{Recall}}{\mathrm{Precision}+\mathrm{Recall}}
\end{align*}
with TP, FP, FN corresponding to True Positive, False Positive, and False Negative rates. Therefore, the maximum score in Trojan-free and Trojan-implanted netlists were 2 and 3, respectively.

A runtime requirement of 30 minutes per testcase was imposed, assuming evaluation inside a Docker image on an AWS Ubuntu~22.04 host (4$\times$ Xeon~8259CL vCPUs, 16\,GB RAM, 1$\times$~Tesla~T4 16\,GB, Nvidia~535.183.01, Docker~26.1.3 with NVIDIA Container Toolkit). The use of multi-threading/GPU was allowed.

\vspace{-1mm}
\subsection{Comparison of Results}

Table \ref{tab:trojan_detection_results} shows comparison of LoRD with BERT in Trojan-implanted netlists. The Trojans are grouped in four types based on functionality \cite{ICCAD2025-contest}. The netlist IDs corresponding to each Trojan type is listed in the last column, as identified by LoRD. (The contest did not require predicting the Trojan type but recommended adding this requirement in future editions.) For each row in the table, we report the average of the evaluation metrics across all corresponding netlist IDs (given in the last column). This is due to lack of space, having 40 Trojan-implanted testcases. At the same time, we note the entries in the Score column in LoRD's case are all very close to the maximum score of 3, with an average of 2.957. 

The BERT-based approach was developed with four BERT models trained, one for each Trojan group, as discussed in Section \ref{sec:BERT}. We ran each BERT model on the corresponding netlist IDs listed per Trojan group. This is the reason a Score column is not reported for BERT-based approach as it is only used for localization (and not Trojan detection). As can be seen, the F1 score in BERT is on-average 0.254 which is much lower than the average F1 score of 0.957 in LoRD. Interestingly, the highest F1 score in BERT is for Trojans T8 and T9 which were from industrial datapaths.  In the BERT-based approach, \texttt{Recall} was always significantly higher than \texttt{Precision}, which indicates a general bias toward predicting that a gate belongs to a Trojan. This behavior was observed among all contestants, as mentioned in the contest final presentation \cite{ICCAD2025-contest}. In general, considering transformer-based models such as \cite{NtNDet_2025} for Trojan detection, their reported high accuracy metrics may be related to a restricted validation, as noted in \cite{krieg2023trusthub}. 

We attribute the lower performance of BERT to several factors: 
(i) a \emph{label granularity mismatch}—the contest task is fundamentally a \emph{set} (boundary) identification problem, but the model is asked to decide gate-by-gate; 
(ii) \emph{boundary-gate semantics}—Trojan “signature” is concentrated near the boundary between trigger/payload and host logic, whereas our tokenization (fan-in cones truncated to a fixed depth) biases the model toward local structure and obscures boundary cues; 
and (iii) \emph{limited supervision}—the available labeled designs are too few and structurally diverse (post-synthesis variability), which hurts generalization.
In contrast, LoRD encodes boundary-aware rules derived from the public RTL definitions and organizers' clarifications, which appears to align better with the contest’s scoring objective (F1 on the \emph{set} of Trojan gates).

        \begin{table}
  \centering
  \caption{Comparison with top-five teams in the contest.}
  \footnotesize
  \setlength{\tabcolsep}{6pt}
  \begin{tabular}{c|ccc|c}
    \toprule
    \scriptsize
    \multirow{2}{*} &
      \multicolumn{3}{c|}{\textbf{Trojan-implanted netlists}} &
      \multicolumn{1}{c}{\textbf{Trojan-free netlists}} \\
    & \textbf{Precision} & \textbf{Recall} & \textbf{F1} & \textbf{Classification Accuracy} \\
    \midrule
    \textbf{LoRD}   & 0.965 & 0.963 & 0.957 & 95\% \\
    \textbf{Top 5 teams}  & 0.861 & 0.915 & 0.886 & 70\% \\
    \bottomrule
  \end{tabular}\label{tab:top5}\vspace{-4mm}
\end{table}

Table \ref{tab:top5} shows comparison of LoRD with top five contest teams. For Trojan-implanted cases, the contest only provided the average precision, recall, and F1 score across the top five teams. As can be seen, LoRD has better values for all metrics, with average F1 score of 0.965 (very close to the maximum 1) while the average F1 score of the top five teams is 0.886. 
Table \ref{tab:top5} also shows the evaluation results for Trojan-free netlists. Specifically, in these test cases, LoRD achieves a classification accuracy of 95\%, correctly identifying all 19 out of 20 Trojan-free hidden netlists, significantly outperforming the 70\% classification accuracy of the top five teams. The testcase which was not correctly classified by LoRD was netlist \#43, which according to the contest results, all teams incorrectly identified this netlist to be Trojan-implanted while it was actually Trojan-free. 





LoRD was implemented in Python and is computationally light. It processed the entire 60 hidden test cases in under a minute, far below the 30 minute per testcase limit imposed by the contest. This was with our local machine with Intel Core i7, an NVIDIA GeForce RTX 4070 SUPER (12GB GDDR6X), and 64 GB of DDR5 memory. In contrast, the BERT-based approach imposes hours to train per model on our local machine. For inference, the BERT-based approach took 10-20 seconds per netlist due to tasks such as tokenization and batching over all cell instances in each netlist. 

\section{Conclusions}
This work addressed the challenge of detecting and localizing RTL-inserted hardware Trojans in synthesized gate-level netlists without relying on golden chips. We showed that RTL Trojans exhibit stable structural and signal-flow patterns at gate-level, enabling effective detection based on graph traversal schemes, word detection, and signal flow tracking, rather than relying on handcrafted numerical features that many ML-based schemes use. We introduced LoRD, a lightweight heuristic-driven approach that achieves near-perfect accuracy on ICCAD 2025 contest benchmarks while requiring minimal computational resources. Compared to a BERT-based  baseline and top five teams in the contest, LoRD significantly outperforms in both detection and localization. 
Our evaluation was based on the contest’s synthesis flows and cell libraries; extending LoRD to broader industrial flows is an interesting direction for future work.

\balance
\bibliographystyle{IEEEtran}
\bibliography{ref}

@inproceedings{tree_emb_Shiv_2019,
 author = {Shiv, Vighnesh and Quirk, Chris},
 booktitle = {Neural Information Processing Systems (NIPS)},
 title = {Novel positional encodings to enable tree-based transformers},
 volume = {32},
 year = {2019},
 pages={12081 - 12091}
}

@article{turc2019wellread,
  author    = {Iulia Turc and Ming{-}Wei Chang and Kenton Lee and Kristina Toutanova},
  title     = {Well-Read Students Learn Better: The Impact of Student Initialization on Knowledge Distillation},
  journal   = {CoRR},
  volume    = {abs/1908.08962},
  year      = {2019},
  eprinttype = {arXiv},
}

@inproceedings{krieg2023trusthub,
  title     = {Reflections on Trusting TrustHUB},
  author    = {Christian Krieg},
  booktitle = {International Conference on Computer-Aided Design (ICCAD)},
  pages     = {1--9},
  year      = {2023},
}

@inproceedings{chen2024trojanformer,
  title     = {{TrojanFormer}: Resource-Efficient Hardware Trojan Detection Using Graph Transformer Network},
  author    = {Menghui Chen and Xiaoyong Kou and Gongxuan Zhang},
  booktitle = {IEEE International Conference on Electronic Technology (ICET)},
  pages     = {165--170},
  year      = {2024},
}

@inproceedings{Latibari2025Transformers,
  author    = {Banafsheh Saber Latibari and Najmeh Nazari and Avesta Sasan and Houman Homayoun and Pratik Satam and Soheil Salehi and Hossein Sayadi},
  title     = {Transformers for Secure Hardware Systems: Applications, Challenges, and Outlook},
  booktitle = {Great Lakes Symposium on VLSI (GLSVLSI)},
  pages     = {841--848},
  year      = {2025},}

@article{Yu2021TrojanDetection,
  title     = {Deep Learning-based Hardware Trojan Detection with Block-based Netlist Information Extraction},
  author    = {Shichao Yu and Chongyan Gu and Weiqiang Liu and Maire O'Neill},
  journal   = {IEEE Transactions on Emerging Topics in Computing (TETC)},
  year      = {2021},
  month     = {Oct},
}

@article{touvron2023llama,
  author    = {Hugo Touvron and Thibaut Lavril and Gautier Izacard and Xavier Martinet and Marie-Anne Lachaux and Timothée Lacroix and Baptiste Rozière and Naman Goyal and Eric Hambro and Faisal Azhar and Aurelien Rodriguez and Armand Joulin and Edouard Grave and Guillaume Lample},
  title     = {{LLaMA}: Open and Efficient Foundation Language Models},
  journal   = {arXiv preprint arXiv:2302.13971},
  year      = {2023},
}

@misc{ICCAD2025-contest, title = {{CAD Contest at ICCAD, Problem A}},  
year  = {2025},
note  = {[Online]. Available: ICCAD 2025 CAD Contest website}
}

@INPROCEEDINGS{MLP11Features_IOLTS17,
  author={Hasegawa, Kento and Yanagisawa, Masao and Togawa, Nozomu},
  booktitle={International Symposium on On-Line Testing and Robust System Design (IOLTS)}, 
  title={Hardware {Trojans} Classification for Gate-Level Netlists Using Multi-Layer Neural Networks}, 
  year={2017},
  volume={},
  number={},
  pages={227-232},
}

@INPROCEEDINGS{TrustHub1,
  author={Salmani, Hassan and Tehranipoor, Mohammad and Karri, Ramesh},
  booktitle={International Conference on Computer Design (ICCD)}, 
  title={On Design Vulnerability Analysis and Trust Benchmarks Development}, 
  year={2013},
  volume={},
  number={},
  pages={471-474},
  doi={10.1109/ICCD.2013.6657085}}

@article{TrustHub2,
  title={Benchmarking of Hardware {Trojans} and Maliciously Affected Circuits},
  author={Shakya, Bicky and He, Tony and Salmani, Hassan and Forte, Domenic and Bhunia, Swarup and Tehranipoor, Mark},
  journal={Journal of Hardware and Systems Security},
  volume={1},
  pages={85--102},
  year={2017},
  publisher={Springer}
}

@INPROCEEDINGS{TrojanSAINT_ISCAS23,
  author={Lashen, Hazem and Alrahis, Lilas and Knechtel, Johann and Sinanoglu, Ozgur},
  booktitle={International Symposium on Circuits and Systems (ISCAS)}, 
  title={{TrojanSAINT}: Gate-Level Netlist Sampling-Based Inductive Learning for Hardware Trojan Detection}, 
  year={2023},
  volume={},
  number={},
  pages={1-5},
  doi={10.1109/ISCAS46773.2023.10181403}}

@INPROCEEDINGS{HW2VEC_2021,
  author={Yu, Shih-Yuan and Yasaei, Rozhin and Zhou, Qingrong and Nguyen, Tommy and Al Faruque, Mohammad Abdullah},
  booktitle={International Symposium on Hardware Oriented Security and Trust (HOST)}, 
  title={{HW2VEC}: a Graph Learning Tool for Automating Hardware Security}, 
  year={2021},
  volume={},
  number={},
  pages={13-23},
  doi={10.1109/HOST49136.2021.9702281}}

@INPROCEEDINGS{SVM5Features_IOLTS16,
  author={Hasegawa, Kento and Oya, Masaru and Yanagisawa, Masao and Togawa, Nozomu},
  booktitle={International Symposium on On-Line Testing and Robust System Design (IOLTS)}, 
  title={Hardware {Trojans} classification for gate-level netlists based on machine learning}, 
  year={2016},
  volume={},
  number={},
  pages={203-206},
  doi={10.1109/IOLTS.2016.7604700}}

@ARTICLE{GNN4HT_2023,
  author={Chen, Lihan and Dong, Chen and Wu, Qiaowen and Liu, Ximeng and Guo, Xiaodong and Chen, Zhenyi and Zhang, Hao and Yang, Yang},
  journal={IEEE Transactions on Computer-Aided Design of Integrated Circuits and Systems}, 
  title={{GNN4HT}: A Two-Stage {GNN}-Based Approach for Hardware {Trojan} Multifunctional Classification}, 
  year={2025},
  volume={44},
  number={1},
  pages={172-185},
  doi={10.1109/TCAD.2024.3428469}}

@INPROCEEDINGS{SVM_ICSIP21,
  author={Du, Maofan and Huang, Zhao and Chen, Yin and Li, Liang and Wang, Quan and Liu, Jinhui},
  booktitle={International Conference on Signal and Image Processing (ICSIP)}, 
  title={A {HT} Detection and Diagnosis Method for Gate-level Netlists based on Machine Learning}, 
  year={2021},
  volume={},
  number={},
  pages={1070-1074},
  doi={10.1109/ICSIP52628.2021.9688647}}

@article{BERT_arXiv18,
  author       = {Jacob Devlin and
                  Ming{-}Wei Chang and
                  Kenton Lee and
                  Kristina Toutanova},
  title        = {{BERT:} Pre-training of Deep Bidirectional Transformers for Language
                  Understanding},
  journal      = {CoRR},
  volume       = {abs/1810.04805},
  year         = {2018},
  eprinttype    = {arXiv},
  eprint       = {1810.04805}
}

@ARTICLE{SCA_power_TC13,
  author={Narasimhan, Seetharam and Du, Dongdong and Chakraborty, Rajat Subhra and Paul, Somnath and Wolff, Francis G. and Papachristou, Christos A. and Roy, Kaushik and Bhunia, Swarup},
  journal={IEEE Transactions on Computers}, 
  title={Hardware {Trojan} Detection by Multiple-Parameter Side-Channel Analysis}, 
  year={2013},
  volume={62},
  number={11},
  pages={2183-2195},
  doi={10.1109/TC.2012.200}}

@INPROCEEDINGS{SCA_tmperature_ICCAD13,
  author={Forte, Domenic and Bao, Chongxi and Srivastava, Ankur},
  booktitle={International Conference on Computer-Aided Design (ICCAD)}, 
  title={Temperature tracking: An innovative run-time approach for hardware {Trojan} detection}, 
  year={2013},
  volume={},
  number={},
  pages={532-539},
  doi={10.1109/ICCAD.2013.6691167}}

@INPROCEEDINGS{SCA_delay_ICCAD16,
  author={Ismari, D. and Plusquellic, J. and Lamech, C. and Bhunia, S. and Saqib, F.},
  booktitle={International Conference on Computer-Aided Design (ICCAD)}, 
  title={On detecting delay anomalies introduced by hardware {Trojans}}, 
  year={2016},
  volume={},
  number={},
  pages={1-7},
  doi={10.1145/2966986.2967061}}

@INPROCEEDINGS{SCA_delay_HOST15,
  author={Exurville, Ingrid and Zussa, Loie and Rigaud, Jean-Baptiste and Robisson, Bruno},
  booktitle={International Symposium on Hardware Oriented Security and Trust (HOST)}, 
  title={Resilient hardware {Trojans} detection based on path delay measurements}, 
  year={2015},
  volume={},
  number={},
  pages={151-156},
  doi={10.1109/HST.2015.7140254}}

@ARTICLE{HT_intro_2010,
  author={Karri, Ramesh and Rajendran, Jeyavijayan and Rosenfeld, Kurt and Tehranipoor, Mohammad},
  journal={Computer}, 
  title={Trustworthy Hardware: Identifying and Classifying Hardware {Trojans}}, 
  year={2010},
  volume={43},
  number={10},
  pages={39-46},
  doi={10.1109/MC.2010.299}}

@misc{wordpiece_arXiv21,
      title={Fast {WordPiece} Tokenization}, 
      author={Xinying Song and Alex Salcianu and Yang Song and Dave Dopson and Denny Zhou},
      year={2021},
      eprint={2012.15524},
      archivePrefix={arXiv},
      primaryClass={cs.CL},
      url={https://arxiv.org/abs/2012.15524}, 
}

@ARTICLE{HTSurvey_2010,
  author={Tehranipoor, Mohammad and Koushanfar, Farinaz},
  journal={{IEEE Design \& Test of Computers}}, 
  title={A Survey of Hardware {Trojan} Taxonomy and Detection}, 
  year={2010},
  volume={27},
  number={1},
  pages={10-25},
  doi={10.1109/MDT.2010.7}}

@article{NtNDet_2025,
title = {{NtNDet}: Hardware Trojan detection based on pre-trained language models},
journal = {Expert Systems with Applications},
volume = {271},
pages = {126666},
year = {2025},
issn = {0957-4174},
doi = {https://doi.org/10.1016/j.eswa.2025.126666},
author = {Shijie Kuang and Zhe Quan and Guoqi Xie and Xiaomin Cai and Xiaoqian Chen and Keqin Li},
}

\end{document}